\def\aa{{A\&A}}

\def\annrev{{ARA\&A}}
\def\apj{{ApJ}}

\documentclass[cup5b]{caps}

\usepackage{graphicx}

\begin{document}

\pagenumbering{arabic}

\author[]{MARIO HAMUY\\The Observatories of the Carnegie Institution of 
Washington}

\chapter{The Latest Version of the Standardized Candle Method for Type II Supernovae}

\begin{abstract}

I use the largest available sample of Type II plateau supernovae to examine
the previously reported luminosity-velocity relation. This study confirms
such relation which permits one to standardize the luminosities of these objects
from a spectroscopic measurement of their envelope velocities, and use them
as extragalactic distance indicators. The ``standard candle method'' (SCM)
yields a Hubble diagram with a dispersion of 0.3 mag, which implies that the SCM produces
distances with a precision of 15\%. Using two nearby supernovae with Cepheid distances I find 
$H_0$=81$\pm$10 km~s$^{-1}$~Mpc$^{-1}$, which compares with 
$H_0$=74 derived from Type Ia supernovae.

\end{abstract}

\section{Introduction}

Type II supernovae (SNe~II, hereafter) are exploding stars characterized by
strong hydrogen spectral lines and their proximity to star forming
regions, presumably resulting from the gravitational collapse of
the cores of massive stars ($M_{ZAMS}$$>$8 $M_\odot$).
SNe~II display great variations in their spectra and lightcurves
depending on the properties of their progenitors at the time
of core collapse and the density of the medium in which they explode (Hamuy 2003a).
The plateau subclass (SNe~IIP) constitutes a well-defined family which can
be distinguished by
1) a characteristic ``plateau'' lightcurve (Barbon et al. 1979),
2) Balmer lines exhibiting broad P-Cygni profiles, and
3) low radio emission (Weiler et al. 2002). These SNe are thought 
to have red supergiant progenitors that do not experience significant mass loss 
and are able to retain most of their H-rich envelopes before explosion.

Although SNe~IIP display a wide range in luminosity, rendering their use as
standard candles difficult, Hamuy \& Pinto (2002) (HP02, hereafter) recently used a
sample of 17 SNe~II to show that their plateau luminosities are well
correlated with the expansion velocities of their ejecta, which implies
that the luminosities can be standardized from a spectroscopic measurement
of the SN ejecta velocity. This ``standard candle method'' (SCM) affords
a new opportunity to derive independent and potentially precise extragalactic distances.
In this paper I use a larger sample of 24 SNe~IIP to assess the precision of the SCM 
and solve for the Hubble constant based on two SNe with Cepheid distances.

\section{The Luminosity-Velocity Relation }

I collected published and unpublished photometric and spectroscopic
data for the largest possible sample of SNe~IIP in order to re-evaluate
the luminosity-velocity relation. The current sample comprises 24 SNe~IIP,
15 of which were included in the HP02 list. Here I exclude SN~1987A and SN~2000cb
from the HP02 sample for exhibiting non-plateau lightcurves.

I measured apparent magnitudes 50 days after explosion (nearly the middle of
the plateau phase) for the 24 SNe~IIP, I applied corrections to the observed
fluxes using my best estimate for dust extinction (see next section), and
converted the fluxes into luminosities from redshift-based distances corrected for the peculiar
motion of the Galaxy and the SN host galaxy (using the flow model described
in the next section). Also, I measured ejecta velocities from
the minimum of the Fe II $\lambda$5169 lines, and I interpolated
them to the same epoch (50 days after explosion) by fitting
a power-law to the observed velocities following the precepts 
described by Hamuy 2001.

\begin{figure}[ht]
\centering
\includegraphics[height=75mm,width=75mm]{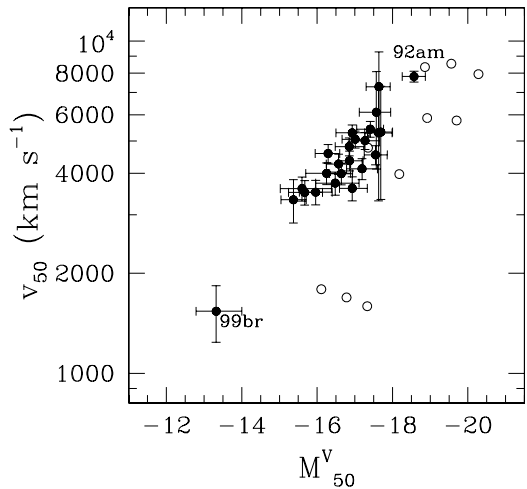}
\caption{Expansion velocities from Fe II $\lambda$5169 versus absolute $V$ 
magnitude, both measured in the middle of the plateau (day 50) of 24 SNe~IIP
(filled circles) and the models of LN83 and LN85 with $\geq$ 8 $M_\odot$
(open circles). }
\label{L_v.fig}
\end{figure}

Fig. \ref{L_v.fig} compares the plateau $V$ luminosity and
ejecta velocity for these 24 SNe~IIP (filled circles)
which reveals the well-known fact that SNe~IIP encompass a
wide range ($\sim$5 mag) in luminosities, and confirms the 
luminosity-velocity relation previously reported by HP02.
The extreme objects in this diagram are
the dim and low-velocity SN~1999br and the luminous
high-velocity SN~1992am. This result reflects the fact that
while the explosion energy increases, so do the kinetic
energy and internal energies. Also plotted in this figure with open circles
are the models of Litvinova \& Nadezhin (1983, 1985) (hereafter LN83 and LN85)
for SNe with progenitors masses $\geq$ 8 $M_\odot$, which reveals
a reasonable agreement with observations.

\section{Determination of Dust Extinction }

In order to use astronomical objects as standard candles
it proves necessary to correct the observed fluxes for dust absorption. 
The estimate of Galactic extinction is under good control 
thanks to the IR dust maps of Schlegel et al. (1998), which permit one to
estimate $A_{GAL}(V)$ to $\pm$0.06 mag. The determination of absorption in
the host galaxy, on the other hand, is not so straightforward.

Here I explore a method which assumes that SNe~IIP should all reach the
same color toward the end of the plateau phase, so a measurement of the
color should give directly the color excess due to dust absorption.
The underlying assumption is that the opacity in SNe~IIP is dominated by e$^-$ scattering,
so they should all reach the same temperature of hydrogen recombination
toward the end of the plateau phase (Eastman et al. 1996).

Next I proceed to test this method using the best-studied plateau SN~1999em
as the reference for the intrinsic color.  For this purpose I adopt the $A_{host}(V)$=0.18
value derived by Baron et al. (2000) from detailed theoretical modeling of the spectra of SN~1999em.
For the other SNe I use their $B-V$ and $V-I$ color at the end of the plateau phase to determine
the color offset relative to SN~1999em and the corresponding visual extinction, $A_{host}(V)$,
assuming the extinction law of Cardelli et al. (1989) for $R_V$=3.1.

\begin{figure}[ht]
\centering
\includegraphics[height=75mm,width=75mm]{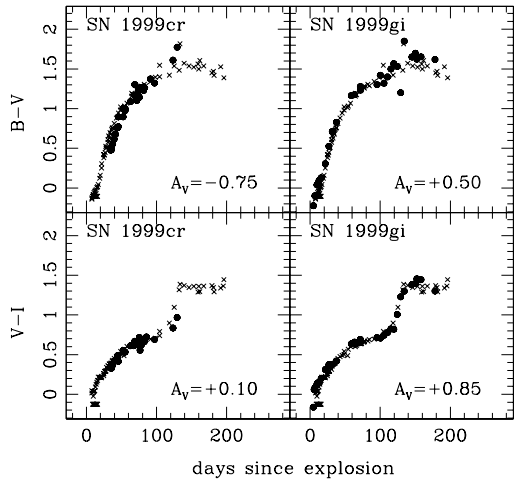}
\caption{(Top) With filled circles are shown the
$B-V$ color curves of SN~1999cr (left) and SN~1999gi (right)
corrected for Galactic extinction, and after applying a color
offset in order to match the dereddened color curves of SN~1999em (crosses).
The required offset for each SN is indicated in each panel in units
of visual extinction $A_V$. (Bottom) Same as above but
for $V-I$. This figure shows that, while the offsets required for
SN~1999gi agree quite well and probably reflect dust extinction
in the host galaxy, the required offsets for SN~1999cr reveal a
poor agreement.}
\label{color.fig}
\end{figure}

The technique is illustrated in Figure \ref{color.fig}. The top panels
show with filled circles the $B-V$ color curves of SN~1999cr (left)
and SN~1999gi (right) corrected for Galactic extinction, after applying an
offset in order to match the dereddened color curves of SN~1999em (crosses).
The required offset for each SN is indicated in units
of visual extinction $A_V$. The bottom panels show the same procedure
using $V-I$ colors.

\begin{table}[ht]
\caption{Galactic and Host-galaxy Extinction for 24 Type II Supernovae.}
\begin{tabular}{lcccc}
\hline \hline
SN & $A_{GAL}(V)$ & $A_{host}(V)$ & $A_{host}(V)$ & $A_{host}(V)$  \\
   & $(\pm$0.06)  & $(B-V)$       & $(V-I)$       & $(\pm0.3)$    \\
\hline
1968L   &  0.219  & -0.90  &   ...      &  0.00  \\
1969L   &  0.205  & -0.70  &   ...      &  0.00  \\
1970G   &  0.028  & -1.20  &   ...      &  0.00  \\
1973R   &  0.107  &  1.40  &   ...      &  1.40  \\
1986I   &  0.129  &  ...   &   0.20     &  0.20  \\
1986L   &  0.099  &  0.30  &   ...      &  0.30  \\
1988A   &  0.136  & -0.40  &   ...      &  0.00  \\
1989L   &  0.123  & -0.60  &   0.90     &  0.15  \\
1990E   &  0.082  &  1.00  &   1.90     &  1.45  \\
1990K   &  0.047  &  0.05  &   0.35     &  0.20  \\
1991al  &  0.168  & -0.30  &   0.10     &  0.00  \\
1991G   &  0.065  &  ...   &   0.00     &  0.00  \\
1992H   &  0.054  &  0.00  &   ...      &  0.00  \\
1992af  &  0.171  & -0.40  &  -0.20     &  0.00  \\
1992am  &  0.164  &  0.35  &   0.20     &  0.28  \\
1992ba  &  0.193  & -0.15  &   0.15     &  0.00  \\
1993A   &  0.572  &  0.00  &   0.10     &  0.05  \\
1993S   &  0.054  &  1.00  &   0.40     &  0.70  \\
1999br  &  0.078  &  0.50  &   0.80     &  0.65  \\
1999ca  &  0.361  &  0.85  &   0.50     &  0.68  \\
1999cr  &  0.324  & -0.75  &   0.10     &  0.00  \\
1999eg  &  0.388  & -0.15  &   0.05     &  0.00  \\
1999em  &  0.130  &  0.18  &   0.18     &  0.18  \\
1999gi  &  0.055  &  0.50  &   0.85     &  0.68  \\
\hline \hline
\end{tabular}
\label{Av.tab}
\end{table}

Table \ref{Av.tab} gives the results for all 24 SNe.
An inspection of this table reveals that the $B-V$ method has serious problems since in
10 cases it yields negative reddenings. This is particularly  pronounced among the historical SNe,
reaching $A_{host}(V)$=-1.2 for SN~1970G. It is possible that part of the problem is due to
inadequate transformations of the photographic magnitudes into the standard Johnson system,
or to background contamination by the host galaxy. However, even SN~1999cr (with modern CCD photometry)
yields a negative value of $A_{host}(V)$=-0.75 (Fig. \ref{color.fig}), which is well beyond
the photometric errors. The $V-I$ method produces independent reddenings for 17 SNe.
This method is much well behaved: only SN~1992af yields a modest negative reddening of $A_{host}(V)$=-0.2.

\begin{figure}[ht]
\centering
\includegraphics[height=75mm,width=75mm]{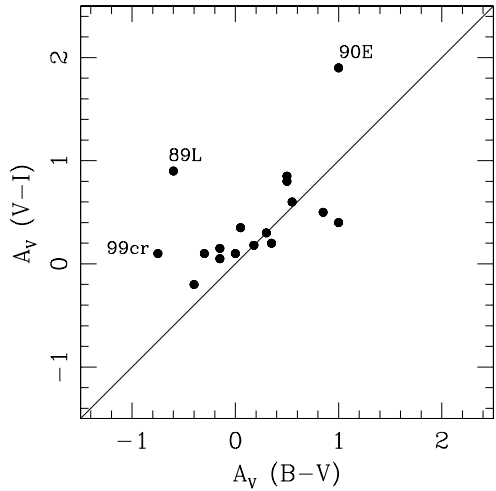}
\caption{Comparison between host-galaxy visual absorption extinction ($A_V$)
of SNe~IIP from $B-V$ and $V-I$ color curves. The ridge
line is the 45$^\circ$ line and not a fit to the data.}
\label{Av.fig}
\end{figure}

Figure \ref{Av.fig} compares the $A_{host}(V)$ values obtained from $B-V$ and $V-I$ which
reveals serious discrepancies between both methods. It is possible that the problem
is due to metallicity effects which are stronger in the $B$ band where line blanketing
is more pronounced (Eastman et al. 1996, Baron et al. 2003).
In any case, this is an unsatisfactory situation and other techniques 
should be explored in the future.

Although the $V-I$ method should be preferred owing to the smaller sensitivity
to metallicity effects, in what follows I use the average of the $B-V$ and $V-I$ extinction values 
(or the single-color value when only one color is available) since I do not have $V-I$
colors for all SNe. The resulting $A_{host}(V)$ values are listed in Table \ref{Av.tab} to
which I assign an uncertainty of $\pm$0.3 mag based
on the reddening difference yielded by both methods.

\section{The Hubble Diagram}

In a uniform and isotropic Universe we expect locally a linear relation
between distance and redshift. A perfect standard candle located at
different distances should describe a straight line in the magnitude-log($z$)
Hubble diagram, so the observed scatter is a measure of how standard 
the candle is, which is what I proceed to assess for the SCM.

\begin{table}[ht]
\caption{Redshifts, Magnitudes, and Ejecta Velocities of the 24 Type II Supernovae.}
\begin{tabular}{lcccccc}
\hline \hline
SN & $v_{CMB}$           & Redshift        &   $V_{50}$ & $I_{50}$      & $v_{50}$      \\
   &                      & Source$^a$      &            &               &               \\
   &($\pm$187 km s$^{-1}$)&                 &            &               & (km s$^{-1}$) \\
\hline
1968L   &  321 &  1 & 12.03(08) &  ...      & 4020(300)   \\
1969L   &  784 &  1 & 13.35(06) &  ...      & 4841(300)   \\
1970G   &  580 &  2 & 12.10(15) &  ...      & 5041(300)   \\
1973R   &  808 &  2 & 14.56(05) &  ...      & 5092(300)   \\
1986I   & 1333 &  1 & 14.55(20) & 14.05(09) & 3623(300)   \\
1986L   & 1466 &  3 & 14.57(05) &  ...      & 4150(300)   \\
1988A   & 1332 &  1 & 15.00(05) &  ...      & 4613(300)   \\
1989L   & 1332 &  3 & 15.47(05) & 14.54(05) & 3529(300)   \\
1990E   & 1426 &  3 & 15.90(20) & 14.56(20) & 5324(300)   \\
1990K   & 1818 &  3 & 14.50(20) & 13.90(05) & 6142(2000)  \\
1991al  & 4484 &  4 & 16.62(05) & 16.16(05) & 7330(2000)  \\
1991G   & 1152 &  1 & 15.53(07) & 15.05(09) & 3347(500)   \\
1992H   & 2305 &  3 & 14.99(04) &  ...      & 5463(300)   \\
1992af  & 5438 &  4 & 17.06(20) & 16.56(20) & 5322(2000)  \\
1992am  &14009 &  4 & 18.44(05) & 17.99(05) & 7868(300)   \\
1992ba  & 1192 &  3 & 15.43(05) & 14.76(05) & 3523(300)   \\
1993A   & 8933 &  4 & 19.64(05) & 18.89(05) & 4290(300)   \\
1993S   & 9649 &  4 & 18.96(05) & 18.25(05) & 4569(300)   \\
1999br  &  848 &  3 & 17.58(05) & 16.71(05) & 1545(300)   \\
1999ca  & 3105 &  4 & 16.65(05) & 15.77(05) & 5353(2000)  \\
1999cr  & 6376 &  4 & 18.33(05) & 17.63(05) & 4389(300)   \\
1999eg  & 6494 &  4 & 18.65(05) & 17.94(05) & 4012(300)   \\
1999em  &  838 &  3 & 13.98(05) & 13.35(05) & 3757(300)   \\
1999gi  &  706 &  3 & 14.91(05) & 13.98(05) & 3617(300)   \\
\hline \hline
\end{tabular}
$^a$Code:\\
1: ${\bf v_{CMB}}$=$H_0~{\bf D_{SBF}}$, where ${\bf D_{SBF}}$ is SBF distance and $H_0$=78.4.\\
2: ${\bf v_{CMB}}$=$H_0~{\bf D_{Ceph}}$, where ${\bf D_{Ceph}}$ is Cepheid distance and $H_0$=78.4.\\
3: CMB redshift corrected  for peculiar flow model of Tonry et al. (2000).\\
4: Heliocentric redshift corrected to CMB.
\label{SN.tab}
\end{table}

For the 8 most distant SNe ($cz$$>$3000 km~s$^{-1}$) I used the observed heliocentric
redshifts of their host galaxies, either from the NASA/IPAC Extragalactic Database or my
own measurement (Hamuy 2001), and converted them to the Cosmic Microwave Background (CMB)
frame in order to remove the observer's peculiar motion. In these cases I neglected the
peculiar motion of the hosts since they are small compared to their cosmological redshifts.
In the remaining cases this effect is potentially significant so I attempted to correct it
using the parametric model for peculiar flows of Tonry et al. (2000). In this model
the CMB velocity of a galaxy is given by

\begin{equation}
{\bf v_{CMB}} = H_0~{\bf D_{SBF}} + {\bf v_{pec}},
\label{pm_eq}
\end{equation}

\noindent where ${\bf D_{SBF}}$ is the SBF distance of Tonry et al. (2001)
in the Ferrarese et al. (2000) scale (F00), ${\bf v_{pec}}$ is the peculiar
velocity of the galaxy which is a function of ${\bf D_{SBF}}$ and galactic
coordinates, and $H_0$ is the Hubble constant which has a value of 78.4 km~s$^{-1}$~Mpc$^{-1}$.

For 9 galaxies I used ${\bf v_{CMB}}$ and inverted equation \ref{pm_eq} to
solve for ${\bf D_{SBF}}$, from which the velocity component free of
peculiar motion, $H_0~{\bf D_{SBF}}$, could be trivially computed.
In 5 cases it was possible to assign the SN host to a galaxy group
and use the corresponding SBF distance given by Tonry et al. (2001)
to compute $H_0~{\bf D_{SBF}}$ directly. In two
cases I used Cepheid distances in the F00 scale. The resulting redshifts
are listed in Table \ref{SN.tab} for the 24 SNe. In all cases I assigned
an uncertainty of $\pm$187 km~s$^{-1}$, which  corresponds to the cosmic
thermal velocity yielded by the model for peculiar flows.

The other ingredients for the Hubble diagram are the apparent magnitudes 
and the ejecta velocities of the SNe. A convenient choice (but not the only one)
is to use magnitudes in the middle of the plateau, so I interpolated the
observed $V$ and $I$ fluxes to the time corresponding to 50 days after explosion.
Other  parameterizations might yield better results and should be explored in the future.
The ejecta velocities come from the minimum of the Fe II $\lambda$5169
lines interpolated to day 50 (as described in Hamuy 2001), which is good
to $\pm$300 km~s$^{-1}$. In 4 cases it was necessary to extrapolate velocities
so I adopted a greater uncertainty of $\pm$2000 km~s$^{-1}$. The magnitudes
and velocities are given in Table \ref{SN.tab} for all SNe.

\begin{figure}[ht]
\centering
\includegraphics[height=75mm,width=75mm]{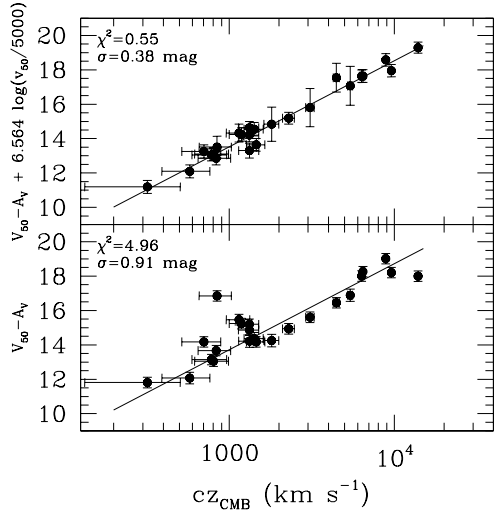}
\caption{(bottom) Raw Hubble diagram from SNe~II plateau $V$ magnitudes.
(top) Hubble diagram from $V$ magnitudes corrected for envelope expansion
velocities. }
\label{hd3.fig}
\end{figure}

The bottom panel of Fig. \ref{hd3.fig} shows the Hubble diagram in the $V$ band,
after correcting the apparent magnitudes of Table \ref{SN.tab} for the reddening
values in Table \ref{Av.tab}, while the top panel shows the same magnitudes
after correction for expansion velocities. A least-squares fit to the data in the
top panel yields the following solution,

\begin{equation}
V_{50} - A_{V} + 6.564(\pm0.88)~log (v_{50}/5000) = 5~log(cz) - 1.478(\pm0.11).
\label{veqn_1}
\end{equation}

\noindent The scatter drops from 0.91 mag to 0.38 mag, thus demonstrating that
the correction for ejecta velocities standardizes the luminosities of SNe~IIP
significantly. It is interesting to note that part of the spread comes from
the nearby SNe which are potentially more affected by peculiar motions of their
host galaxies. When the sample is restricted to the eight objects with 
$cz$$>$3,000 km s$^{-1}$, the scatter drops to only 0.33 mag.
The corresponding fit for the restricted sample is,

\begin{equation}
V_{50} - A_{V} + 6.249(\pm1.35)~log (v_{50}/5000) = 5~log(cz) - 1.464(\pm0.15).
\label{veqn_2}
\end{equation}

\begin{figure}[ht]
\centering
\includegraphics[height=75mm,width=75mm]{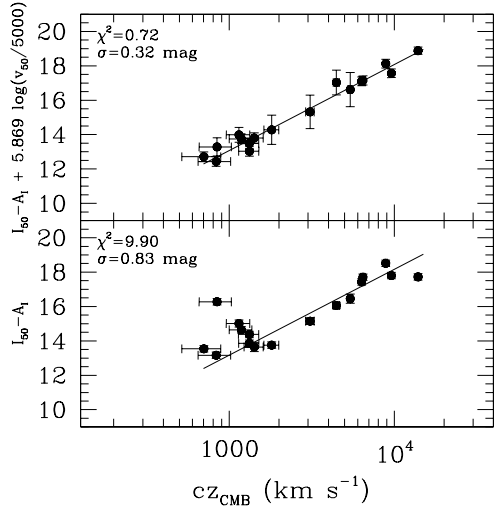}
\caption{(bottom) Raw Hubble diagram from SNe~II plateau $I$ magnitudes.
(top) Hubble diagram from $I$ magnitudes corrected for envelope expansion
velocities. }
\label{hd4.fig}
\end{figure}

Figure \ref{hd4.fig} shows the same analysis but in the $I$ band. In this case the
scatter in the raw Hubble diagram is 0.83 mag, which drops to 0.32 mag
after correction for ejecta velocities. This is even smaller that the 0.38
spread in the $V$ band, possibly due to the fact that the effects of dust
extinction are smaller at these wavelengths. The least-squares fit yields the
following solution,

\begin{equation}
I_{50} - A_{I} + 5.869(\pm0.68)~log (v_{50}/5000) = 5~log(cz) - 1.926(\pm0.09).
\label{ieqn_1}
\end{equation}

When the eight most distant objects are employed the spread is 0.29 mag,
similar to that obtained from the $V$ magnitudes and  the same sample, and the solution is,

\begin{equation}
I_{50} - A_{I} + 5.445(\pm0.91)~log (v_{50}/5000) = 5~log(cz) - 1.923(\pm0.11).
\label{ieqn_2}
\end{equation}

This analysis shows that the SCM can produce relative distances with a precision of 15\%,
but more objects in the Hubble flow are required to pin down the actual
precision of this technique.

\section{The Value of the Hubble Constant}

The SCM can be used to solve for the Hubble constant, provided
a distance calibrator is available. If the distance $D$ of the calibrator
is known, the Hubble constant is given by

\begin{equation}
H_0(V) = \frac {10^{V_{50} - A_V + 6.249 log(v_{50}/5000) + 1.464}}{D},
\label{h0_v}
\end{equation}

\noindent when the distant sample is adopted.  A similar expression applies to the $I$ band data.

Among the objects of our sample SN~1968L, SN~1970G, SN~1973R, and SN~1999em have
precise Cepheid distances, but only two of them have been published so far
(Freedman et al. 2001). The distances and the corresponding $H_0$ values are
summarized in Table \ref{H0.tab}. 
Within the uncertainties the two estimates are in good agreement, and the average
proves to be 81$\pm$10 km~s$^{-1}$~Mpc$^{-1}$. This result compares with
the 74 value derived from Type Ia SNe calibrated in the Freedman et al. (2001) scale
(Phillips et al. 2003), which lends further credibility to the SCM.

\begin{table}[ht]
\caption{The Hubble Constant.}
\begin{tabular}{lccc}
\hline \hline
SN & Distance &  $H_0(V)$                 \\
   & Modulus  &  (km~s$^{-1}$~Mpc$^{-1}$) \\
\hline
1970G   & 29.13(11)    & 77$\pm$13         \\
1973R   & 29.86(08)    & 87$\pm$15         \\
\hline
Average &              & 81$\pm$10    \\
\hline \hline
\end{tabular}
\label{H0.tab}
\end{table}

HP02 found a value of $H_0$=55$\pm$12 based on one calibrator (SN~1987A),
which proves significantly lower than the current 81$\pm$10 value. The main reason
for this difference is that SN~1987A is not a plateau event and should not have been
included in the HP02 sample since the physics of its lightcurve is different
than that of SNe~IIP. Unlike these objects which have red supergiants 
progenitors (Arnett 1996, Hamuy 2003b), SN~1987A exploded as a compact blue
supergiant (Woosley et al. 1987) and most of the energy deposited in the
envelope by the shock wave formed after core collapse went into adiabatic expansion, thus leading to
a dimmer plateau and to a lightcurve promptly powered by 
$^{56}$Ni $\rightarrow$ $^{56}$Co $\rightarrow$ $^{56}$Fe (Blinnikov et al. 2000).

The Cepheid distances for SN~1968L and SN~1999em will be released soon
(G. Tammann and D. Leonard, priv. comm.), thus improving significantly the $H_0$
value from the SCM. SN~1999em will prove particularly important because 
it has far better observations than any of the other 3 calibrators,
both in the $V$ and $I$ bands.

\section{Conclusions}

I used the largest possible sample of SNe~IIP to examine their use as
standardized candles. The main conclusions of this study are the following,

\noindent 1) The luminosity-velocity relation previously reported by HP02
is confirmed from this larger sample of 24 SNe.

\noindent 2) The luminosity-velocity relation can be used to standardize the luminosities
of SNe~IIP, and the corresponding Hubble diagram has a dispersion of 0.3 mag,
which implies that SNe~IIP can produce distances with a precision of 15\%.

\noindent 3) Using two nearby SNe with Cepheid distances I find a value of $H_0$=81$\pm$10,
which compares with the 74 value derived from Type Ia SNe.

\noindent 4) The least satisfactory aspect of the SCM is the dereddening method based
on the SN color curves, which produces different results depending on the
color used. Clearly other techniques need to be explored in the future.

\noindent 4) This study confirms that SNe~IIP offer a great potential as distance
indicators. These conclusions, however, are based on several historical SNe with
poor photometric observations and few objects in the quiet Hubble flow.
Clearly a greater sample of SNe with $cz$$>$3000 km~s$^{-1}$ and
modern CCD photometry is needed.
The recently launched Carnegie Supernova Program at Las Campanas Observatory has
already targeted a dozen such SNe and in the next two years it will
produce an unprecedented database of spectroscopy and optical/infrared photometry
for a large number of SNe, which will allow us to achieve this goal.

\noindent 5) Although the precision of the SCM is only half as good as that produced by 
SNe~Ia, with the 8-m class telescopes currently in operation it should be possible 
to get spectroscopy of SNe~IIP down to $V$$\sim$23 and start populating the Hubble diagram
up to $z$$\sim$0.3. A handful of SNe~IIP will allow us to check the distances to SNe~Ia.

\vspace{0.05in}

I dedicate this work to my dear friends and collaborators Bob Schommer
and Marina Wischnjewsky, for their tireless work which led to an
enormous progress of the supernova field over recent years.
Support for this work was provided by NASA through Hubble Fellowship
grant HST-HF-01139.01-A awarded by the Space Telescope Science Institute,
which is operated by the Association of Universities for Research in Astronomy,
Inc., for NASA, under contract NAS 5-26555.

\end{document}